\documentclass[pra,preprint,twocolumn,10pt,amsmath,amssymb,nofootinbib,bm,bbm,notitlepage,superscriptaddress]{revtex4-1}

\usepackage[english]{babel}
\usepackage[utf8]{inputenc}
\usepackage[T1]{fontenc}

\setcitestyle{sort&compress,numbers}
\usepackage{amsthm, color, mathtools, amsmath, amssymb, setspace,bbm, tensor, subfigure, mathtools, placeins, graphicx, wrapfig,float, verbatim,enumitem}
\usepackage[dvipsnames]{xcolor}
\usepackage[unicode=true, breaklinks=false, pdfborder={0 0 1}, backref=false, colorlinks=true, linkcolor=blue, citecolor=blue]{hyperref}
\usepackage{cancel,bm}
\usepackage{bbm}
\usepackage{orcidlink}
\usepackage{amsbsy}

\let\oldsqrt\sqrt
\def\sqrt{\mathpalette\DHLhksqrt}
\def\DHLhksqrt#1#2{%
\setbox0=\hbox{$#1\oldsqrt{#2\,}$}\dimen0=\ht0
\advance\dimen0-0.2\ht0
\setbox2=\hbox{\vrule height\ht0 depth -\dimen0}%
{\box0\lower0.4pt\box2}}

\usepackage[normalem]{ulem}

\def\bra#1{\mathinner{\langle{#1}|}}

\def\ket#1{\mathinner{|{#1}\rangle}}

\def\braket#1{\mathinner{\langle{#1}\rangle}}

\newcommand*\xbar[1]{%
   \hbox{%
     \vbox{%
       \hrule height 0.5pt 
       \kern0.5ex
       \hbox{%
         \kern-0.2em
         \ensuremath{#1}%
         \kern-0.0em
       }%
     }%
   }%
} 

\def\BraVert{\egroup\,\mid\,\bgroup}

\def\ketbra#1#2{\ket{#1\vphantom{#2}}\!\bra{#2\vphantom{#1}}}

\def\bra#1{\mathinner{\langle{#1}|}}

\def\ket#1{\mathinner{|{#1}\rangle}}

\def\braket#1{\mathinner{\langle{#1}\rangle}}

\usepackage{bbm, braket}
\usepackage[mathscr]{euscript}
\allowdisplaybreaks
\newtheorem*{theorem*}{Theorem}

\newtheorem*{corollary*}{Corollary}

\newtheorem*{observation*}{Observation}

\newcommand{\nn}{\nonumber}
\newcommand{\cv}[1]{(#1)}
\newcommand{\cvb}[1]{[#1]}
\newcommand{\cvc}[1]{\{#1\}}
\newcommand{\cvv}[1]{\vert #1\vert}

\newcommand{\oper}[1]{\hat{#1}}

\newcommand{\ds}{\displaystyle}

\begin{document}

\title{A Simple Scheme for Preparation of the Approximate Gottesman-Kitaev-Preskill States with Random Walk Mechanism}
\author{Fattah Sakuldee\,\orcidlink{0000-0001-8756-7904}}
\email{fattah.sakuldee@ug.edu.pl}
\affiliation{Wilczek Quantum Center, School of Physics and Astronomy, Shanghai Jiao Tong University, 800 Dongchuan Road, Minhang District, Shanghai, 200240, China}
\affiliation{The International Centre for Theory of Quantum Technologies, University of Gda\'nsk, Jana Ba\.zy\'nskiego 1A, 80-309 Gda\'nsk, Poland}

\begin{abstract}
The Gottesman-Kitaev-Preskill (GKP) coding is proven to be a good candidate for encoding a qubit on continuous variables (CV) since it is robust under random-shift disturbance. Its preparation in optical systems, however, is challenging to realize in nowadays state-of-the-art experiments. In this article, we propose a simple optical setup for preparing the approximate GKP states by employing a random walk mechanism. We demonstrate this idea by considering the encoding on the transverse position of a single-mode pulse laser. We also discuss generalization and translation to other types of physical CV systems.
\end{abstract}

\date{\today}
		
\maketitle

\section{Introduction}
Manipulation over continuous variables (CV) to execute quantum features in light or other optical systems is one of the promising research aspects in quantum information science  \cite{Serafini2017,Braunstein2005}. Its potential applications have been discussed in several areas ranging from a quantum key distribution  \cite{Denys2021,Kanitschar2022,Primaatmaja2022,Srikara2020}, entanglement and resources theory \cite{Srikara2020,Simon2000,Duan2000,Hyllus2006,Mihaescu2020,Kumar2021,Bowen2004}, quantum metrology and states discrimination \cite{Notarnicola2023,Lvovsky2009,Smithey1993,Wang2020,Wu2021}, to quantum communication and state transfers \cite{Dequal2021,Rani2023,Notarnicola2023,Lin2020,Bose2023,Anuradha2023,Peuntinger2014}. 
One of the advantages of this program is the ability to encode a qubit into continuous degrees of freedom that are intact to disturbances from the environment, by only using traditional sources of quantum systems and coherence, especially light. Two examples are Gottesman-Kitaev-Preskill (GKP) encoding \cite{Gottesman2001} and cat encoding \cite{Cochrane1999}. The former is robust against a small shift in either position or momentum space while the latter is intact under photon losses \cite{Fabre2020A}. 

The GKP code, in particular, is an interesting encoding scheme since gates, error correction, and measurements can be done within the spirit Gaussian formalism where any operators are degree two polynomials in creation and annihilation operators of the bosonic field \cite{Hastrup2021A}. This has opened up several analyses of GKP states in various aspects of quantum computation science. 
Error correction is discussed for both computation problems \cite{Li2023,Zhang2023,Siegele2023,Zhang2021,Wan2020,Grimsmo2021,Wang2019,Vuillot2019,Raveendran2022,Noh2022,Bangar2023} and communication problems \cite{Schmidt2023,Fabre2023,Lin2023,Fukui2021,Hastrup2021A}. 
A similar issue on noise analysis and noise suppression is examined in detail in Refs.~\cite{Hanggli2020,Noh2019,Hastrup2021C,Tzitrin2020,Rosati2018}. 
Several works demonstrate characterisations of the GKP states \cite{Walshe2023,Stafford2022,Shaw2022,Conrad2022,Seshadreesan2022,Shi2019}, manipulations of the GKP states for some applications \cite{Rani2023,Rojkov2023,Baranes2023,Yamasaki2020a,Yamasaki2020b,Fukui2023}, or the tomography concerning the GKP states \cite{Wu2023,Hahn2022,Hastrup2021B}.

Another challenge involves the preparation of the GKP state in natural physical systems, which aligns with the primary objective of this article. 
Most recent experimental demonstrations are performed using a microwave cavity field coupled to a superconducting circuit \cite{CampagneIbarcq2020}, and using the motional state of a trapped ion \cite{Flhmann2019}. For optical systems, the experimental realization is not well established despite the fact that there are several proposals in the literature. The suggested techniques include the use of superposition of optical coherent states with post-selection \cite{Vasconcelos2010}, and without post-selection \cite{Weigand2018}; the coupling of an atomic ensemble to a squeezed state of light \cite{Motes2017}; the heralding of by the measurement on few modes of multimode Gaussian states \cite{Su2019,Eaton2019}; the strong interactions of free electrons with light \cite{Dahan2023}; the engineering a time-periodic Hamiltonian whose Floquet states are GKP states \cite{Kolesnikow2023}; dissipation of Harmonic oscillators \cite{Sellem2023B}; photon counting assisted node-teleportation \cite{Eaton2022}; a cavity QED system \cite{Hastrup2022}; a microwave frequency comb parametrically modulating a Josephson circuit \cite{Sellem2023A}; a cross-Kerr interaction between a squeezed light and a superposition of Fock states \cite{Fukui2022}; and the time-frequency continuous degrees together with spontaneous parametric down conversion \cite{Fabre2020B}.

All the proposals mentioned above, despite their potential implementations, are still challenging in actual experiments. Also, many of them involve state-of-the-art optics to prepare the GKP states. In this work, we show that such states can be prepared using only simple optical linear elements. The overall procedure can be inscribed as a random walk on phase space, in which the polarization of the light plays the role of coin-states. We propose an experiment setup adapted from a modified Sagnac module introduced in Ref.~\cite{Walborn2018}. In that work, they use a variable slanted mirror to generate a shift in momentum space corresponding to a transverse position of a light beam.
In our case, we use the same module, with iteration, to implement a symmetric walk on the momentum space. We also show that the use of a rotation induces a similar walk scheme in the conjugate position space. With this principle, one can generate a Binomial distribution on either momentum or position line in the phase space, providing a promising approximation to the GKP states.

In principle, our proposed scheme is similar in spirit to the mechanism proposed in Ref.~\cite{Vasconcelos2010}, e.g. inducing a superposition of coherence states by post-selection, however, our ready-to-use description for the preparation procedure has its own novelty. 
Also, it can be seen that the same principle can be applied to time-frequency degrees of freedom or solid-state-based memory registers.

The article is organized as follows. The mathematical formalism used in this paper includes the relevant description of GKP states given in Sec.~\ref{sec:prelim}. We review, in Sec.~\ref{sec:random-walk}, a random walk mechanism and relevant operation on the phase space. Later, we translate the mechanism into linear optical setups and then discuss a procedure for constructing GKP states in Sec.~\ref{sec:optical-setup}. Finally, the discussions and conclusions are given in Sec.~\ref{sec:conclusion}.

\section{Preliminary}\label{sec:prelim}
In this section, we recall a formalism of the continuous variable system and its phase space, and later we discuss the characteristics of GKP states. We focus on a one-dimensional system while the extension to a higher dimension will be seen as straightforward.

\subsection{Mathematical Framework}\label{sec:math}
Let $\mathcal{H}$ be a Hilbert space of square-integrable complex-valued functions on a real line, i.e., \begin{equation}
\mathcal{H}=L^2\cv{\mathbb{R}}=\cvc{\psi:\mathbb{R}\rightarrow\mathbb{C}: \int_{\mathbb{R}}\cvv{\psi\cv{x}}^2dx < \infty}.
\end{equation}
Define a position operator $\oper{x}$ and a momentum operator $\oper{p}$ on $\mathcal{H}$ as multiplicative operators $\oper{x}\psi\cv{x}{=}x\psi\cv{x}$ and derivative operators $\oper{p}\psi\cv{x}{=}-i\partial_x\psi\cv{x}$ respectively. 
The position and momentum operators follow the canonical commutation relation $\cvb{\oper{x},\oper{p}}:=\oper{x}\oper{p}-\oper{p}\oper{x}=i,$ where $i$ is the unit of imaginary numbers. 
The Hilbert space $\mathcal{H}$ can be spanned by two sets of bases; a continuous one $\cvc{\delta\cv{x}: x\in\mathbb{R}}$ and a discrete set of states $\cvc{\phi_n\cv{x}:n=0,1,\ldots},$ where $\oper{H}_{\rm{har}}\phi_n\cv{x}=\cv{n+\frac{1}{2}}\phi_n\cv{x}$ defines the eigenstates of a harmonic oscillator $\oper{H}_{\text{har}}=\frac{1}{2}(\oper{x}^2+\oper{p}^2).$ The former representation indeed leads to the name \emph{continuous variables.} 

For the discrete basis, one can define an annihilation operator $\oper{a}$ and a creation operator $\oper{a}^\dagger$ by $\oper{a}=\frac{\oper{x}+i\oper{p}}{\sqrt{2}},$ where one can show that $\oper{a}\phi_n=\sqrt{n}\phi_{n-1}$ for $ n>0$ and $\oper{a}\phi_0=0$ (a vacuum state) and $\oper{a}^\dagger$ is its Hermitian conjugate. The canonical commutation will then be $\cvb{\oper{a},\oper{a}^\dagger}=1.$ We also use a Dirac notation for the states and the dual states, i.e. $\ket{\psi}$ represents the function $\psi$ and $\ket{n}$ represents $\phi_n$ for $n=0,1,\ldots.$ Also we write $\ket{x}$ and $\ket{p}$ eigenfunctions of $\oper{x}$ and $\oper{p}$ respectively.

A displacement operator $\oper{D}$ is defined as an operator associated with a translation of the position or momentum variable, i.e. $\oper{D}\cv{x,p}=e^{-ix\oper{p}+ip\oper{x}},$ or in a single complex argument $\oper{D}\cv{\alpha}=\exp\cv{\alpha\oper{a}^\dagger - \overline{\alpha}\oper{a}}$ where $\alpha=\frac{x+ip}{\sqrt{2}}.$ Note that the displacement parameter $\alpha,$ is a complex number where the real part corresponds to \textit{displace} in position space and the imaginary part is to \textit{displace} in momentum space.  

In this work, we are interested in a superposition state $\rho$ of coherent states, which is described by
    \begin{equation}
		\ket{\psi} = \sum_{r,s}c_{r,s}\ket{r,s}\label{eq:rho-def},
    \end{equation}
where $\sum_{r,s}\cvv{c_{r,s}}^2=1,$ and 
    \begin{align}
    		\ket{r,s} &= \ds\int_{-\infty}^\infty dx \psi_{\sigma_{r,s}}\cv{x-r}\ket{x} \label{eq:Gaussian-element},\\
    		\psi_{\sigma_{r,s}}\left(x-r\right) &= \left(\frac{1}{2\pi\sigma_{r,s}^2}\right)^{1/4}e^{-\tfrac{\cv{x-r}^2}{4\sigma_{r,s}^2}+isx}, \label{eq:Gaussian-element-psi}
    \end{align}
with $(r,s)$ indicating the center of the Gaussian state $\rho_{r,s}$ on the phase space, equipped with spreading parameter $\sigma_{r,s}^2,$ in which we assume from now on for simplicity. 
With these indices one can write a vacuum state as
	\begin{align}
		\ket{vac} &:= \ket{0,0}= \frac{1}{\sqrt{2\pi\sigma^2}}\int_{-\infty}^{\infty} dx \psi_{\sigma}\cv{x}\ket{x} \label{eq:vacuum-ket},
	\end{align}
when we assume that $\sigma^2_{r,s}=\sigma^2$ for all $r$ and $s.$ It is easy to see that the state in Eq.~\eqref{eq:rho-def} can be transcribed into a combination of displacement operators acting on the vacuum state, i.e.,
	\begin{align}
		\ket{\psi} &=  \Bigg[\sum_{r,s}c_{r,s}\oper{D}\cv{r,s}\Bigg]\ket{vac}  \label{eq:rho-mix-in-D}.
	\end{align}
In other words, one can prepare an arbitrary superposition of coherent states (of equal variance) by preparing a vacuum state and a superposition of displacement operators. The latter can be technically prepared by controlled operations with the preparation or the measurement in a superposition state on the controlling ancilla. This essence can be directly applied to the GKP states as we can see in the following.

\subsection{A GKP Encryption}\label{sec:GKP-encryption}
The $d-$dimensional Gottesman-Kitaev-Preskill states are qudit states encoded in phase space using several orthogonal functions and effective Pauli manipulations upon them \cite{Gottesman2001}. 
The ideal case is to use a Dirac distribution $\delta(x-x')=\braket{x\vert x'}$ centered at various points as canonical basis elements and to construct the GKP states by taking $d$ different summations over them. For example, for the qubit case, one of the computational bases \cite{Gottesman2001} consists of
	\begin{align}
		\ket{\overline{0}} &= \sum_{r=-\infty}^{\infty}\ket{x=2rw} \label{eq:GKP-0},\\
		\ket{\overline{1}} &= \sum_{r=-\infty}^{\infty}\ket{x=(1+2r)w} \label{eq:GKP-1}.
	\end{align}
for some positive number $w.$ The summations above are employed to make the states resistant to the displacement noise, with a strength comparatively smaller than $w$---the half-width between two adjacent teeth. 
From now on we will consider only GKP qubits for simplicity, while we can see later that the generalization to a higher dimension can be easily obtained within the same mechanism with slight modifications.

However, as discussed in the original paper Ref.~\cite{Gottesman2001}, such states are not normalized and the Dirac distribution is not physically accessible. In practice, approximate GKP states should be considered by replacing the Dirac distribution with a collection of coherent functions $\ket{r,s}$ defined in Eq.~\eqref{eq:Gaussian-element} and introducing a Gaussian weight function in the basis elements. Namely, suppose that $\sigma^2_{r,0}=\sigma^2 $ for all $r,$ one may approximate states in Eqs.~\eqref{eq:GKP-0} and \eqref{eq:GKP-1} by
	\begin{align}
		\ket{\widetilde{0}} &= N_0\sum_{r=-\infty}^{\infty}e^{-\frac{1}{2}\kappa^2(2rw)^2}\ket{2rw,0} \label{eq:GKP-Gauss-0},\\
		\ket{\widetilde{1}} &= N_1\sum_{r=-\infty}^{\infty}e^{-\frac{1}{2}\kappa^2\cvb{(1+2r)w}^2}\ket{(1+2r)w,0} \label{eq:GKP-Gauss-1}.
	\end{align}
where $\kappa$ is a positive number that denotes the width of the envelope, and $N_0$ and $N_1$ are normalization factors.  
One can see that these are special cases of superposition of coherence states of the form Eq.~\eqref{eq:rho-def}. Furthermore, such states can be inscribed as combinations of displacement operations in a vacuum state (with variance $\sigma^2$) in the spirit of Eq.~\eqref{eq:rho-mix-in-D}. To complete the formulation, in the following section, we review another relevant ingredient---the random walk on phase space.

\section{Random Walk Mechanism}\label{sec:random-walk}
Here we discuss the random-walk mechanism on phase space. First, we discuss a random walk scheme in position direction, i.e., the direction of variable $x$ when the walk is generated by a momentum operator $\oper{p}.$ Second, we discuss a similar scheme in momentum direction, when these two primary schemes can be related by a $\frac{\pi}{2}$ rotation.

\subsection{Random Walk on Position Space}\label{sec:generic-construction}
A random walk mechanism describing a sequence of changes of a target state, called a walker state, by given transition operators, with conditions determined by an additional state called the coin state. 
In our problem, we can define a walker state from the distribution on the phase space as follows. Let us consider a one-dimensional lattice with separation distance $2w$ embedded in the real line corresponding to position variable $x.$ In this sense the distributions $\delta\cv{x-2rw}$ for $r\in\mathbb{Z}$ represent the canonical walker states occupying the positions $2rw.$ 
For a binary walk, the transition operators can be described by displacement operators $\oper{D}\cv{\pm 2w}$ where the signs represent the directions of the walk. 

For the coin state, we introduce another degree of freedom attached to the walker state. 
We consider the simplest coin state, a qubit, and we will call it a polarization state with basis elements $\ket{H}$ and $\ket{V},$ representing horizontal and vertical polarizations, respectively. 
A single-step walk can be written as a controlled unitary operator on a walker-coin composite state as
	\begin{equation}
		\oper{U} = \oper{D}\cv{w}\otimes\ketbra{H}{H} + \oper{D}\cv{-w}\otimes\ketbra{V}{V} \label{eq:def-walker-U}.
	\end{equation}
The directional preference is incorporated by preparation and post-selection of the coin state. Note that the quantum coherence remains by the mechanism above, but we will not use the term quantum random walk since it is not relevant to our problem.  
For the $N$-step walk, it can be done by repeating the procedure for $N$ times.

\subsection{GKP States in Terms of Random Walk}\label{sec:GKP-walk}
Now consider the situation when we prepare the initial state in $\ket{\psi}\otimes\ket{+},$ where $\ket{\pm}=\frac{\ket{H}\pm\ket{V}}{\sqrt{2}},$
following by an $2N-$repetition of walk operator Eq.~\eqref{eq:def-walker-U} interlacing with projection onto polarization state $\ket{+}$ and post-selecting in the same polarization state $\ket{+}.$ 
The reduced operator on the walker state reads $\cvb{\oper{D}\cv{w}+\oper{D}\cv{-w}}^{2N},$ which transforms the walker state $\ket{\psi}$ to a state with Bernoulli distribution
    \begin{equation}
        \ket{\psi_+\cv{2N}} = \dfrac{1}{2^{2N}}\sum_{r=-N}^N \left(\begin{array}{c} 2N \\ N-r \end{array}\right) \oper{D}\cv{2rw}\ket{\psi} \label{eq:psi-N},
    \end{equation}
and when the number of steps is large enough, i.e., $N\gg 1,$ it can be approximated to 
    \begin{equation}
        \ket{\psi_+\cv{2N}} \approx \dfrac{1}{\sqrt{\pi N}}\sum_{r=-N}^N e^{-r^2/N}\oper{D}\cv{2rw}\ket{\psi} \label{eq:psi-N-large}.
    \end{equation}
It is clear at this point that by setting $\ket{\psi}=\ket{vac},$ the random walk-induced states in both Eqs.~\eqref{eq:psi-N} and \eqref{eq:psi-N-large} are approaching an approximate GKP state $\ket{\overline{0}}.$ Furthermore, it is a truncation of the state $\ket{\widetilde{0}}$ in Eq.~\eqref{eq:GKP-Gauss-0} with $\kappa=(2Nw^2)^{-1/2}$ and $N_0=(\pi N)^{-1/2}.$ This exemplifies that it is possible to use a random walk scheme to prepare an approximate GKP state.

For the state $\ket{\overline{1}},$ one can apply a similar procedure with a slight modification. 
For example, one can either set the initial state to $\ket{\psi}=\oper{D}\cv{w}\ket{vac},$ or apply an additional displacement operator $\oper{D}\cv{w}$ to the output $\ket{\psi_+\cv{2N}}.$ Hence it suffices to demonstrate only the construction Eqs.~\eqref{eq:psi-N} and \eqref{eq:psi-N-large}.

\subsection{Random Walk on Momentum Space}\label{sec:momentum-walk}
In the previous construction, we consider a random walk on the lattice embedded in the spatial part of the phase space. A similar mechanism can be done also on the momentum line. 
Technically, it can be implemented by replacing the walker operators $\oper{D}\cv{\pm w}$ in Eq.~\eqref{eq:def-walker-U} by $\oper{D}\cv{\pm iw},$ generating a momentum kick on the input walker state $\ket{\psi}.$ For $2N$ steps walk with an initial state $\ket{\varphi}\otimes\ket{+}$
   \begin{align}
       \ket{\varphi_+\cv{2N}} &= \dfrac{1}{2^{2N}}\sum_{r=-N}^N \left(\begin{array}{c} 2N \\ N-r \end{array}\right) \oper{D}\cv{2irw}\ket{\varphi} \label{eq:phi-N-momentum},\\
        &\approx \dfrac{1}{\sqrt{\pi N}}\sum_{r=-N}^N e^{-r^2/N}\oper{D}\cv{2irw}\ket{\varphi} \label{eq:phi-N-large-momentum},
   \end{align}
for large enough $N.$ For the vacuum initial state $\ket{\varphi}=\ket{vac},$ the resulting state does not directly approximate the state $\ket{\overline{0}}$ but rather a rotated state $\ket{\overline{0}_R}:=\oper{R}_{\frac{\pi}{2}}\ket{\overline{0}},$ where $\oper{R}_{\frac{\pi}{2}}$ is a $\tfrac{\pi}{2}$ rotation $\oper{R}_{\frac{\pi}{2}}=e^{-i\pi\cv{\oper{x}^2+\oper{p}^2}/4}$ on the phase space.

One can further observe that, not only the resulting states $\ket{\overline{0}}$ and $\ket{\overline{0}_R}$ can be related via $\oper{R}_{\frac{\pi}{2}},$ but also do the whole random walk schemes in position space
Eq.~\eqref{eq:psi-N} and momentum space Eq.~\eqref{eq:phi-N-momentum}. Namely, we have $\oper{D}\cv{\pm iw}=\oper{R}_{\frac{\pi}{2}}\oper{D}\cv{\pm w}\oper{R}^\dagger_{\frac{\pi}{2}},$ and $\ket{\varphi_+\cv{2N}}=\oper{R}_{\frac{\pi}{2}}\ket{\psi_+\cv{2N}}$ where $\ket{\varphi}=\oper{R}_{\frac{\pi}{2}}\ket{\psi}.$ 
In other words, there is no significant difference between the random walk schemes on position or momentum space. The rotation operator $\oper{R}_{\frac{\pi}{2}}$ employed in this equivalence is simply the operator realization of Fourier transform on phase space, which can be implemented by known optical elements e.g., lenses arrays. In the next section, we are demonstrating on an optical setup the random walk-induced construction of an approximate GKP states  Eqs.~\eqref{eq:phi-N-momentum} and \eqref{eq:phi-N-large-momentum} on the momentum space, on an optical setup, and later we will show that it can be modulated to obtain the states Eqs.~\eqref{eq:psi-N} and \eqref{eq:psi-N-large} in terms of position variable.

\section{Optical Setups for Random Walk and Construction of the Approximate GKP States}\label{sec:optical-setup}
Here we demonstrate the implementation of the random walk-induced construction of the approximate GKP states on an optical setup. Note that, though we employ the basic elements and use a simplified model of bulk optics, one can easily translate them to other similar optical setups with higher fidelity, e.g., using photonic media, fiber optical elements or active elements for beam splitting.

\subsection{Simple Optical Setup}\label{sec:RW-setup}
\begin{figure}[!h]
    \centering
        \includegraphics[width=0.95\columnwidth]{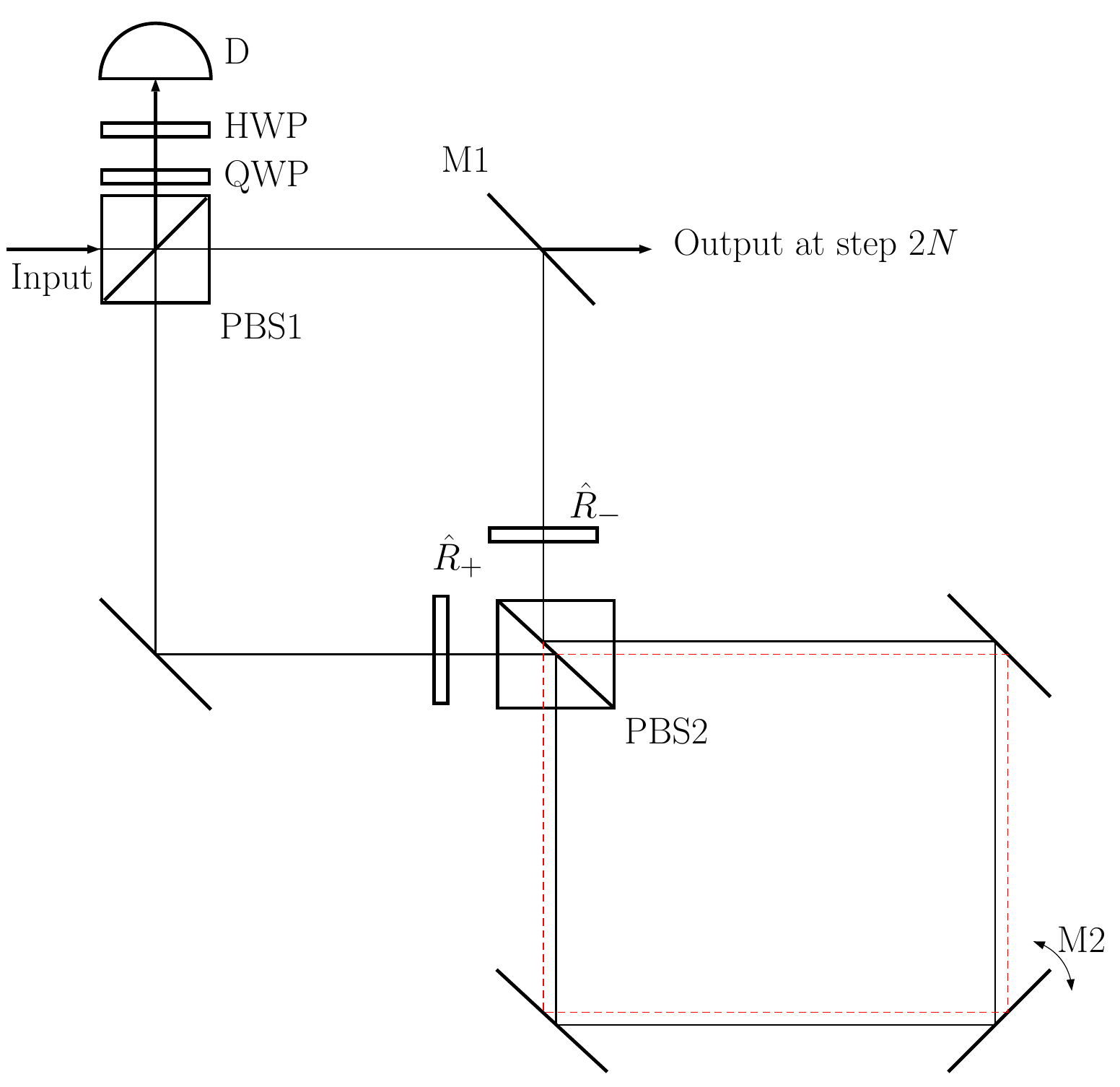}
    \caption{A simple schematic of the walking module: An input state is prepared in $\ket{vac}\otimes\ket{V}$ and passed to the interferometer. After passing the polarizing beam splitter PBS$1,$ the mirror M$1,$ which is set to be a perfect mirror, will reflect the beam to the polarization rotation operator $\oper{R}_-$ creating a superposition in polarization state. The beam is then split by PBS$2$ into two paths, each experiencing a different induced misalignment by the mirror M$2,$ and the two split beams recombine and pass to the polarization rotation operator $\oper{R}_+.$ The latter creates a superposition of the split paths for each tagged polarization state, which will be post-selected by PBS$1$ where the symmetric superposition state will pass to the next iteration. After $2N$ repetitions, the mirror M$1$ is turned to be perfectly transparent, through which the state passing at the output port is propositional to the target state $\ket{\varphi_+\cv{2N}}$ as expected.}\label{fig:modules}
\end{figure}

The main objective of this section is to propose a basic optical setup for the random walk-induced construction of the GKP states. 
First, let us consider the setting depicted in Fig.~\ref{fig:modules}. We consider a laser beam whose polarization exhibits a coin state while the transverse position is the target continuous variable for the GKP encryption. The principle of this interferometer relies on the separated beams with different tagged polarisation, produced by the polarizing beam splitter PBS$2$ traveling in two different directions (black solid line and red dot line in Fig.~\ref{fig:modules}), clockwise and anticlockwise, and experiencing a momentum kick by slant mirror M$2.$ Let the angle of the slanted mirror be $w,$ and assume that it is small enough, e.g. in the order of microradian, the action of the kick on the momentum space associated with the transverse position can be effectively described by $\oper{D}\cv{\pm i w}$ \cite{Walborn2018}, where the sign refers to the direction of the traveling beam. The two paths then recombine and are post-selected by the polarizing beam splitter PBS$1$ leading to a superposition of two shifted states on the spatial part. The first step of the walk will produce a cat state while iterating the walk by $2N$ times will provide our target state $\ket{\varphi_+\cv{2N}}$ in Eq.~\eqref{eq:phi-N-momentum}.

For a particular picture, let us consider the following protocol. First, prepare an input state in $\ket{vac}\otimes\ket{V}.$ The first polarizing beam splitter PBS$1$ is adjusted in such a way that the input state will pass to the interferometer beginning at PBS$2.$ 
The only active element in this setup is the mirror M$1.$ It is a switchable mirror that exhibits either a perfect mirror or a clear glass depending on the applied voltage. We adjust it to be a perfect mirror for the iteration step $1,\ldots,2N-1$ and to be perfectly transparent at the final step $2N.$ After this element we apply an operator $\oper{R}_-:=\ketbra{-}{H}+\ketbra{+}{V}$ on the polarization part, converting the state to $\ket{vac}\otimes\ket{+}.$ The second polarizing beam splitter PBS$2$ will split the beam into two paths, each of which will experience the momentum kick by M$2,$ leading to a state
    \begin{equation}
        \dfrac{\ket{w}\otimes\ket{H}}{\sqrt{2}}+\dfrac{\ket{-w}\otimes\ket{V}}{\sqrt{2}} \label{eq:phi_0-po-M2}.
    \end{equation}
Note the sign in the spatial part can be chosen without loss of generality. 
Now the state will recombine at PBS$2$ (while we discard another path from the beam splitter,) and then its polarization is rotated by $\oper{R}_+=\ketbra{+}{H}+\ketbra{-}{V}.$ The effective state will then read
    \begin{equation}
        \dfrac{\ket{w}\otimes\ket{+}+\ket{-w}\otimes\ket{-}}{\sqrt{2}} \label{eq:phi_0-po-R+}.
    \end{equation}
Finally, after the application of PBS$1$ the state at the detector D for step $1$ becomes 
    \begin{equation}
        \dfrac{1}{\sqrt{2}}\ket{\varphi_-\cv{1}}\otimes\ket{V} :=\dfrac{1}{\sqrt{2}}\dfrac{\ket{w}-\ket{-w}}{\sqrt{2}}\otimes\ket{H} \label{eq:phi_1--at-D},
    \end{equation}
and the reflected state,
    \begin{equation}
       \dfrac{1}{\sqrt{2}}\ket{\varphi_+\cv{1}}\otimes\ket{V} := \dfrac{1}{\sqrt{2}}\dfrac{\ket{w}+\ket{-w}}{\sqrt{2}}\otimes\ket{V} \label{eq:phi_1+},
    \end{equation}
will define the input state for the next step. Note that one can employ the state $\ket{\varphi_-\cv{1}}$ at the detector to infer the information of the target state $\ket{\varphi_+\cv{1}}.$

Perform an iteration by $2N$ times the output state after the switchable mirror M$1$ (which is perfectly transparent at this step) will then read  
    \begin{equation}
        \dfrac{1}{2^{N}}\ket{\varphi_+\cv{2N}}\otimes\ket{V} \label{eq:phi_+2N-at-M1}.
    \end{equation}
By tracing out the polarization part and renormalizing the spatial CV part one can obtain $\ket{\varphi_+\cv{2N}}$ or the approximate GKP state Eq.~\eqref{eq:phi-N-momentum} as claimed.

\subsection{Rotation to the Spatial Walk}\label{sec:rotation-to-spatial}
As previously discussed, the random walk scheme in momentum and position spaces can be related by conjugation of a Fourier operator $\oper{R}_{\frac{\pi}{2}}.$ For the scheme represented above, we have utilized the momentum kick as a walker operator and construct the state $\ket{\varphi_+\cv{2N}}$ to approximate $\ket{\overline{0}_R}.$ Such a mechanism can be transformed into a mechanism to prepare the approximation of the state $\ket{\overline{0}}$ from the vacuum state.

In terms of optical elements, the Fourier transform between spatial position and momentum can be implemented by an ideal concave lens \cite{Stoler1981,Ozaktas1995,Jagoszewski1998}. Technically, for the ideal case, a diverging lens of focal length $f,$ at the distance $f$ after the setting for the momentum walk, and then the state at the image plane at $f$ away from the lens will be $\oper{R}^\dagger_{\frac{\pi}{2}}\ket{\varphi_+\cv{2N}}.$ We assume that the wave vector $k$ of the light is the same as the focal length $f.$ In a more generic scenario, the same principle can be applied to the whole process in Fig.~\ref{fig:modules}, i.e., placing a converging lens before the momentum walk setup and a diverging lens at the other end of the setup, with both appropriate distances, corresponds to the transformation
    \begin{align}
        \frac{1}{2^{2N}}&\sum_{r=-N}^N \left(\begin{array}{c} 2N \\ N-r \end{array}\right) \oper{D}\cv{2irw}\nn\\
        &\Rightarrow  \frac{1}{2^{2N}}\sum_{r=-N}^N \left(\begin{array}{c} 2N \\ N-r \end{array}\right) \oper{D}\cv{2rw} \label{eq:walk-xp-transform}.
    \end{align}
By setting a vacuum state $\ket{\psi}=\ket{vac}$ as the input state (walker), one can construct an approximation for the GKP state $\ket{\overline{0}}$ in the same way as in the momentum walk. 
With a similar modular concept, one can also generate an approximation of the state $\ket{\overline{1}_R}$ by setting an initial state to $\ket{\varphi}=\oper{D}(iw)\ket{vac}$ in the momentum walk scheme, i.e. inducing a misalignment by $w$ from the beam center for the input state in Fig.~\ref{fig:modules}. Also, one can use the same setup to generate an approximation of $\ket{\overline{1}}$ by conjugation of converging and diverging lenses, transforming the momentum random walk to a random walk in position space as discussed.

\section{Discussions and Conclusions}\label{sec:conclusion}
We have discussed so far a protocol for the preparation of approximate GKP states by random walk mechanism. 
We remark that, although the demonstration for the protocol is done by a simple optical setup, the mechanism itself has explicit generalizations in several directions. Let us discuss some of these to elucidate our claim.

\paragraph{Generalization to Grid states:}
One of the direct generalizations is the extension to higher dimensional CV systems, e.g. $2-$dimension CV system. For instance, one can consider displacement operators on the plane associated with the wavefront of the laser beam and utilize the so-called grid states---a qubit GKP states defined on $2-$dimension lattice embedded on the plane orthogonal to the light propagating direction \cite{Sellem2023B}, e.g. 
    \begin{align*}
        \ket{\overline{0}}_2 &:= \sum_{r_x,r_y=-\infty}^{\infty}\ket{x=2r_xw,y=2r_yw}\\
        \ket{\overline{1}}_2 &:= \sum_{r_x,r_y=-\infty}^{\infty}\ket{x=(1+2r_x)w,x=(1+2r_y)w}.
    \end{align*}
The implication for the optical setup is straightforwardly adaptable from the setup in Fig.~\ref{fig:modules} when the mirror M$2$ shall incorporate a slant in two directions. 
Another type of grid states is the states involving a $2$-d lattice on the phase space (of a one-dimensional bosonic system) within the spirit of Eq.~\eqref{eq:rho-mix-in-D}. The optical implementation for the preparation of such states will involve both momentum-walk and position-walk setups introduced in this work. Furthermore, in principle, one can apply both slants and rotation to create hybrid states which are GKP states in both position-momentum and time-frequency degrees of freedom.

\paragraph{Time-frequency degrees of freedom:}
Apart from using the position-momentum degrees of freedom as a CV system, one can instead employ the time-frequency degrees of freedom as the physical framework. In this sense, the displacement operators in position and momentum spaces will correspond to the time translation (unitary evolution) and energy shift respectively \cite{Fabre2020A}. The preparation of the GKP states using time-frequency degrees of freedom as CV systems using linear optical setups, and its related error correction scheme has been discussed fruitfully in Refs.~\cite{Fabre2020B,Fabre2023}. One can observe that our generation protocol can be adopted as an alternative method for the preparation of the time-frequency GKP states. For instance, similar to the protocol in the mentioned references, one can fix the mirror M$2$ with zero slants and add a phase shifter in the loop after PBS$2.$ Another method is fixing the mirror M$2$ and applying a rotation to the whole setup, creating a Sagnac effect that induces advanced and retarded phase shifts for two separated paths. These two modifications simply replace the displacement operator in Eq.~\eqref{eq:def-walker-U} with a phase shift operator, considered as a walker operator on the energy space. In this sense, the output states of the random walk protocol will exhibit the time-frequency GKP states.

\paragraph{Solid state dynamical degrees of freedom:}
In connection with the observation above, the use of time-frequency degrees of freedom as CV systems can be adapted to other systems apart from the bulk optical setting. For instance, in the solid state system, one may consider a spin coupling to a spin bath or bosonic bath. A similar random walk protocol is discussed in Ref.~\cite{Sakuldee2019}, where the spin plays the role of coin state and the spin bath energy spectra and its Fourier conjugate, the decay time, exhibit the frequency-time degrees of freedom. By performing a random walk protocol with post-selection on the spin state, which is analogous to the post-selection on the polarization state in this work, one can encrypt approximate time-frequency GKP states into effective spin cluster spectra. This suggests that one may use this protocol to utilize a spin cluster in the bath as a register for storing a qubit state (up to the coherent time of the cluster) by using only a classical clock and simple optical readout, i.e. the square shape electromagnetic pulses sequences. 

\paragraph{Conclusion:} We have shown that preparation of GKP states can be done using only elementary linear optical elements. We demonstrate a simple optical setup for preparing the approximate GKP states employing a random walk mechanism on the phase space of the transverse position of a single-mode pulse laser, in which its polarization exhibits a controlling degree of freedom. With the preparation of the initial state in a vacuum state and a post-selection on a particular sequence, one can pass the random walk Bernoulli distribution on the momentum line in the phase space, creating approximate GKP states. Several generalizations and translations to other types of physical CV systems are briefly discussed. 
We expect that our protocol provides a cheap method for preparing a robust qubit for near-term quantum computing and communication.

\section*{Acknowledgements}
Comments and suggestions from {\L}ukasz Rudnicki and Ekkarat Phongophas are gratefully appreciated. FS acknowledges support from the Foundation for Polish Science (IRAP project, ICTQT, contract no. 2018/MAB/5, co-financed by EU within Smart Growth Operational Programme).

\bibliographystyle{apsrev4-1}
\bibliography{main.bbl}

\end{document}